\documentclass[12pt]{article}

\usepackage{physics} 
\usepackage{siunitx}
\usepackage{enumerate} 
\usepackage{pgfplots}
\usepackage{pgfplotstable}
\usepackage{tikz,pgfplots}
\usepackage{amsmath}
\usepackage{wasysym}
\usepackage{geometry}
\usepackage{float}
\usepackage{cite}
\usepackage{xcolor}
\usepackage{tabularx}
\usepackage{array}

\newcolumntype{Y}{>{\centering\arraybackslash}X}

\pgfplotsset{compat=1.14}
\begin{document}

\title{Simultaneous photonic and phononic bandgaps in a hexagonal lattice geometry with gradually transforming circular-to-triangular air gap holes}

\author{Suhas Suresh Bharadwaj$^{1}$ and Adarsh Ganesan$^{1*}$ \\
\small $^{1}$Department of Electrical and Electronics Engineering,\\
\small Birla Institute of Technology and Science, Pilani – Dubai Campus,\\
\small Dubai International Academic City, Dubai, UAE, 345055 \\
\small $^*$adarshvganesan@gmail.com}

\maketitle 

\begin{abstract} 
The integration of photonic and phononic bandgaps within a single scalable architecture promises transformative advances in optomechanical and acousto‐optic devices. Here, we design and simulate a two-dimensional hexagonal lattice in silicon with air‐gap holes that transition smoothly from circular to triangular via tuneable geometrical parameters including air-gap hole radius ($R$) and tether length ($l$). By independently varying these two parameters, we systematically explore diverse honeycomb lattice geometries and their bandgap properties. This transformation from circular to triangular air-gap holes enables suppression of both electromagnetic and elastic wave modes through Bragg scattering and symmetry modulation. We demonstrate that systematic variation of \textit{\textit{R}} and \textit{\textit{l}} allows tuning of photonic and phononic bandgaps upto 49.7\% and 24.8\% respectively. This possibility of geometrically tuning bandgaps provide a strong foundation for applications in Bragg filters, sensors etc. without the need for complex defects or exotic materials.
\end{abstract}

\section{Introduction}
	 
Photonic crystals are artificially engineered materials with periodic variations in refractive index, designed to control electromagnetic wave propagation through photonic bandgaps \cite{gangwar2023recent, butt2021recent}. These bandgaps arise from destructive interference of light waves due to Bragg scattering in periodic dielectric structures. Sadat-Saleh et al. \cite{sadat2009tailoring} articulate the periodic modulation of refractive index can prohibit the propagation of electromagnetic waves of a range of different frequencies. The formation of these photonic bandgaps critically depends on lattice geometry, the dielectric contrast of the material, and symmetry, with hexagonal lattices offering unique advantages in mode localization and isotropic light control \cite{cui2024simulation}.\\ 
   
Phononic crystals--like photonic crystals—are also artificially engineered materials with periodic variations in elastic properties, enabling unprecedented control over mechanical wave propagation through phononic bandgaps. These bandgaps arise from destructive interference of elastic waves, suppressing vibrations and thermal phonons in specific frequency ranges. Alternatively, bandgaps may also originate from mechanisms involving strong localized elastic resonances within the phononic crystal \cite{oudich2023tailoring, croenne2011band}. These engineered bandgaps enable wave confinement, filtering, and waveguiding, with applications in optomechanics, mechanical vibration shielding, and energy harvesting \cite{laude2021principles, gonella2009interplay}. Recent advancements highlight their role in thermal management and noise reduction, particularly in GHz-frequency ranges that are critical for room-temperature applications \cite{vasileiadis2021progress}. Phononic crystals can be constructed through a variety of periodic arrangements with different geometrical shapes and materials, and 2D designs using silicon in particular have demonstrated the possibility of bandgap tunability via geometric parameters such as periodicity, lattice-thickness (three-dimensional structures), porosity, etc. \cite{yu2019phononic, anufriev2016reduction}. \\

Over the past three decades, several studies have demonstrated the coexistence of photonic and phononic bandgaps--collectively termed as phoxonic crystals--in a single architecture. Early numerical work by Sadat-Saleh et al. \cite{sadat2009tailoring} showed that photonic and phononic bandgaps can be tailored in membrane‐pillar arrays via careful tuning of pillar spacing and diameter. El Hassouani et al. \cite{el2010dual} extended this concept to a thin‐plate geometry, reporting dual photonic and phononic bandgaps in a periodic array of silicon pillars on a substrate, while Bria et al. \cite{bria2011opening} achieved localized photonic and phononic bandgaps in square‐lattice perforated slabs by exploiting slab thickness as an extra degree of freedom. This phenomenon has enabled transformative applications—ranging from low‐threshold lasers \cite{john1987strong} and slow‐light waveguides \cite{joannopoulos2008molding} in photonics to vibration isolation and thermal management in phononics \cite{armenise2010phononic} by harnessing lattice symmetry, filling fraction, and defect engineering to tailor band structures and mode localization. Subsequent research has also analysed slab-thickness \cite{armenise2010phononic, norris2011silicon} and elastic‐modulus dependencies in square and triangular lattices \cite{armenise2010phononic, norris2011silicon, hou2011slab}.\\

It is also established that topology optimization of the 2D structure maximizes and increases the robustness of the bandgap \cite{jia2018designing, dong2014topology}. Also, Pennec et al. \cite{pennec2019phononic} articulated that features like asymmetric cavities or rotated scatterers within a hexagonal framework can facilitate localization of optical and mechanical modes \cite{pennec2019phononic, djafari2011band}. Collectively, these works along with the works of Yuksel et al. highlight the unique versatility of hexagonal lattices in supporting tuneable bandgaps through geometric and symmetry-based modifications \cite{yuksel2024enhanced}. More recently, Baboly et al. \cite{baboly2013effect} revealed that in a hexagonal lattice of tethered elements, phononic bandgap width can be tuned continuously via tether length, highlighting the potent role of mass redistribution and local symmetry breaking in elastic wave control. While silicon-based phoxonic crystal designs have been successfully demonstrated, a broadly scalable framework that enables simultaneous and continuous tuning of both photonic and phononic bandgaps through simple geometric parameters remains lacking. Specifically, a design that allows for rapid bandgap optimization through independent variation of air gap-hole radius ($R$) and tether length ($l$) in a standard hexagonal lattice, without requiring complex topology optimization or defect engineering, has not yet been established. Recently, Abdurakhmonov et al. \cite{abdurakhmonov2025simultaneous} theoretically investigated the possibility of simultaneous existence of photonic and phononic bandgaps in nanoporous anodic aluminum oxide phoxonic crystals.\\

Here, both photonic and phononic bandgaps \textit{'simultaneously'} appear in the same geometrical configuration without requiring them to occur at identical frequency ranges. When both bandgaps form in a periodic lattice with the same lattice period $a$, the Bragg condition requires $\lambda \sim a$ for both systems. Consequently, the frequency ratio between photonic and phononic bandgaps inherently arises due to the velocity ratio:

\begin{equation}
\frac{f_{\text{photonic}}}{f_{\text{phononic}}} \approx \frac{c_{\text{light}}}{c_{\text{sound}}} \approx 35,000
\end{equation}

The simultaneous existence of phoxonic bandgaps allows the control of both electromagnetic and acoustic waves independently in the same structure, which is particularly useful for applications combining optical and mechanical (optomechanical) functions \cite{gavartin2011optomechanical, kipfstuhl2014modeling, aspelmeyer2014cavity}. This approach has been established in prior work \cite{safavi2010design, shin2015control}, where \textit{simultaneous} consistently refers to the coexistence of bandgaps rather than bandgap overlap.\\

This paper demonstrates the possibility of continuously tuneable dual bandgaps in a 2D silicon hexagonal lattice whose air gaps gradually evolve from circular to triangular by an independent control of geometrical parameters viz. \textit{R}) and \textit{l}. Here, a hexagonal lattice is employed owing to its energy absorption capabilities \cite{sliwa2005phononic}. By tuning \textit{R} and \textit{l}, we achieved degrees of tuning of 49.7\% and 24.8\% for photonic and phononic bandgaps respectively, exceeding prior square and triangular designs \cite{klatt2019phoamtonic}. We further analyze the dependence of bandgap on both \textit{R} and \textit{l}, providing a universal geometric framework for phoxonic crystal design and predictive bandgap engineering for future optomechanical and acousto-optic devices \cite{safavi2010design, aspelmeyer2014cavity}.

\section{Simulation Method}
	
Using COMSOL Multiphysics, we designed a 2D unit cell of a hexagonal crystal lattice and its procedure is depicted in Fig.~1. At each vertex of hexagon (Fig.~1(a)), circular sectors of radius \textit{R} are drawn (Fig.~1(b)). The tethers of half-length \textit{l}/2 are further introduced (Fig.~1(c)). The sectors are then cut out from the hexagon to form a geometrical element (Fig.~1(d)). By tiling multiple such geometrical elements, we construct a 2D lattice (Fig.~1(e)). The unit cell has then be identified such that their edges repeat periodically (Fig.~1(f)).\\

For a 2D hexagonal lattice with lattice constant $a$, the irreducible Brillouin zone is traversed along the high-symmetry path of a rectangular supercell (a $\times$ $\sqrt{3}a$): $\Gamma \rightarrow X \rightarrow \Sigma \rightarrow \Gamma$. The coordinates of high-symmetry points are: $\Gamma$ $(0, 0)$, $X$ $(\pi/a, 0)$, and $\Sigma$ $(2\pi/3a, 2\pi\sqrt{3}/3a)$. The wave vector $\mathbf{k} = (k_x, k_y)$ for the parametric variable $k \in [0, 3]$ is given by:\\

\textbf{For $0 \leq k < 1$ \quad ($\Gamma \rightarrow X$):}
\begin{equation} \label{eq3}
k_x = \frac{k\pi}{a}, \quad \quad k_y = 0
\end{equation}\\

\textbf{For $1 \leq k < 2$ \quad ($X \rightarrow \Sigma$):}

\begin{equation} \label{eq4}
k_x = \frac{\pi(4-k)}{3a}, \quad \quad k_y = \frac{2\pi\sqrt{3}(k-1)}{3a}
\end{equation}\\

\textbf{For $2 \leq k \leq 3$ \quad ($\Sigma \rightarrow \Gamma$):}
\begin{equation} \label{eq5}
k_x = \frac{2\pi(3-k)}{3a}, \quad \quad k_y = \frac{2\pi\sqrt{3}(3-k)}{3a}
\end{equation}

At $k = 0$ and $k = 3$, the vector returns to the $\Gamma$ point, ensuring complete coverage of the irreducible Brillouin zone for band structure calculations.\\

Note: In our COMSOL simulation, we introduce a dimensionless variable \textit{k} (not to be confused with the wave vector magnitude $\mathbf{k}$) that varies from 0 to 3 to trace the high-symmetry path through the irreducible Brillouin zone. This parameter \textit{k} serves as a linear index along the three path segments ($\Gamma \rightarrow X \rightarrow \Sigma \rightarrow \Gamma$), with \textit{k} = 0, 1, 2, 3 corresponding to the high-symmetry points $\Gamma, X, \Sigma, \Gamma$ respectively. The actual wave vector components (\textit{$k_{x}$}, \textit{$k_{y}$}) are functions of this parameter as given in eqns. (2)–(4).

\begin{figure}[H]
\centering
\includegraphics[width=0.8\linewidth]{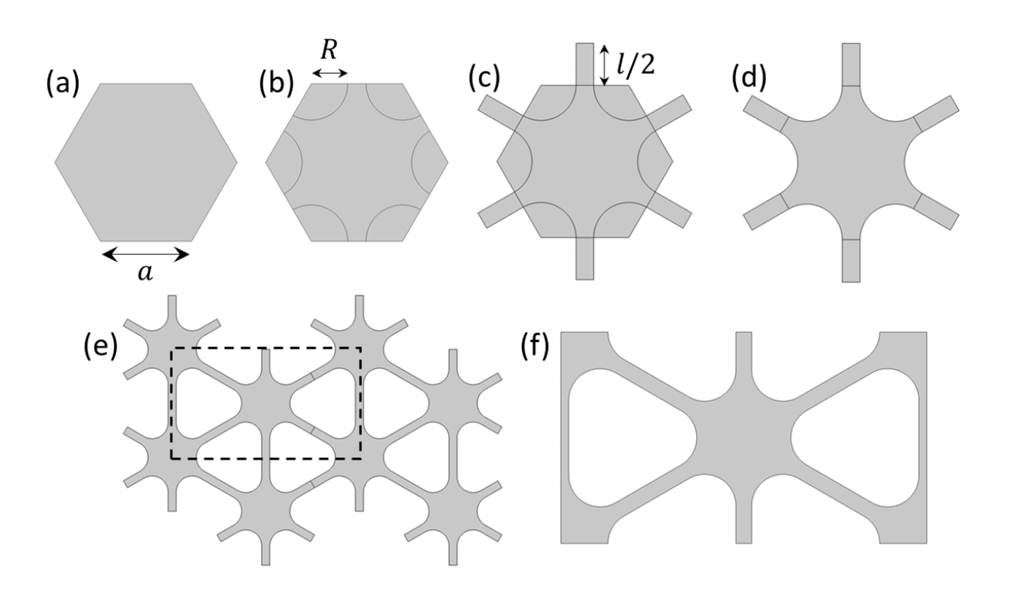}
\caption{Construction of the unit cell. (a) A regular hexagon of side length a; (b) Circular sectors of radius \textit{R} are centered at each vertex of the hexagon; (c) Tethers of length \textit{l}/2 are attached to the hexagon; (d) Removal of the sectors to obtain a geometry element; (e) Tiling the elementary patterns to produce a lattice; (f) The unit cell of the lattice.}
\label{fig1}
\end{figure}

While the material of unit cell is defined as Si (Silicon), the gaps are filled with air. For the simulations of photonic and phononic wave propagation, wave optics and structural mechanics modules in COMSOL Multiphysics are used respectively. The photonic and phononic band structures are computed by solving the Bloch-periodic eigenvalue problem for a unit cell. For photonic waves, the governing equation is:

\begin{equation}
\nabla \times (\frac{1}{n(x)} \times \mathbf{E}(x)) = \lambda_{photonic}~\mathbf{E}(x)
\end{equation}

\noindent where $\lambda_{photonic}$ = $(\frac{\omega_{photonic}}{c})^{2}$, \textbf{E}($x$) is the Electric field, n($x$) is the refractive index and $\omega_{photonic}$ is the angular frequency. \\

\noindent Similarly, for phononic waves, the elastodynamic eigenvalue problem is formulated as:

\begin{equation}
\nabla \cdot (\mathbf{C}(x) : \nabla^{s}\mathbf{u}(x)) +\lambda_{phononic}~\rho(x) \mathbf{u}(x) = 0
\end{equation}

\noindent where $\lambda_{phononic}$ = $(\omega_{phononic})^{2}$, \textbf{C}$(x)$ is the elasticity tensor, \textbf{u}$(x)$ is the Displacement field, $\rho(x)$ is the mass density, and $\nabla^{s}$ is the symmetric gradient operator.\\

\noindent Both equations are solved under their respective Bloch-periodic boundary conditions. For photonic waves, the periodic boundary condition is written as:

\begin{equation}
\mathbf{E}(x + a_{i}) = \mathbf{E}(x) e^{(ik · a_{i})}
\end{equation}

\noindent For phononic waves, the analogous condition applied to the displacement field \textbf{(u$(x)$)} is:

\begin{equation}
\mathbf{u}(x + a_{i}) = \mathbf{u}(x) e^{(ik · a_{i})}
\end{equation}

\noindent where $\mathbf{k}$ = ($k_x, k_y$) corresponds to the wavevector along the irreducible Brillouin zone high-symmetry path defined in eqns. (2)-(4).\\

The photonic and phononic eigenfrequencies are solved by discretizing the Bloch-periodic eigenvalue problem using finite element basis functions and applying the built-in ARPACK eigenvalue solver. For each discrete $k$-point sampled along the high-symmetry path of the irreducible  Brillouin zone, the solver computes the initial N eigenfrequencies. \\

Now, the influence of two key geometrical parameters \textit{l} and \textit{R} on the structure of unit cell is studied. When \textit{l} is increased for a constant radius \textit{R} = 0.48$\,\mu m$, the transformation of air gap holes from being circular to more triangular is observed as shown in Figs. 2(a-e). As \textit{l} increases, the tethers elongate and henceforth the distance between the hexagonal units increases. This results in an apparent transformation of holes from being circular to triangular. Also, by decreasing \textit{R}, the rounded/filleted edges of triangular holes gradually disappear – henceforth resulting in sharp triangular holes (Figs. 2 (f-j)). By tuning \textit{l} and \textit{R}, the holes of lattice can be transformed from circular to triangular and vice versa. Henceforth, the influence of the geometry of holes on the photonic and phononic bandgaps is studied.

\begin{figure}[H]
\centering
\includegraphics[width=1\linewidth]{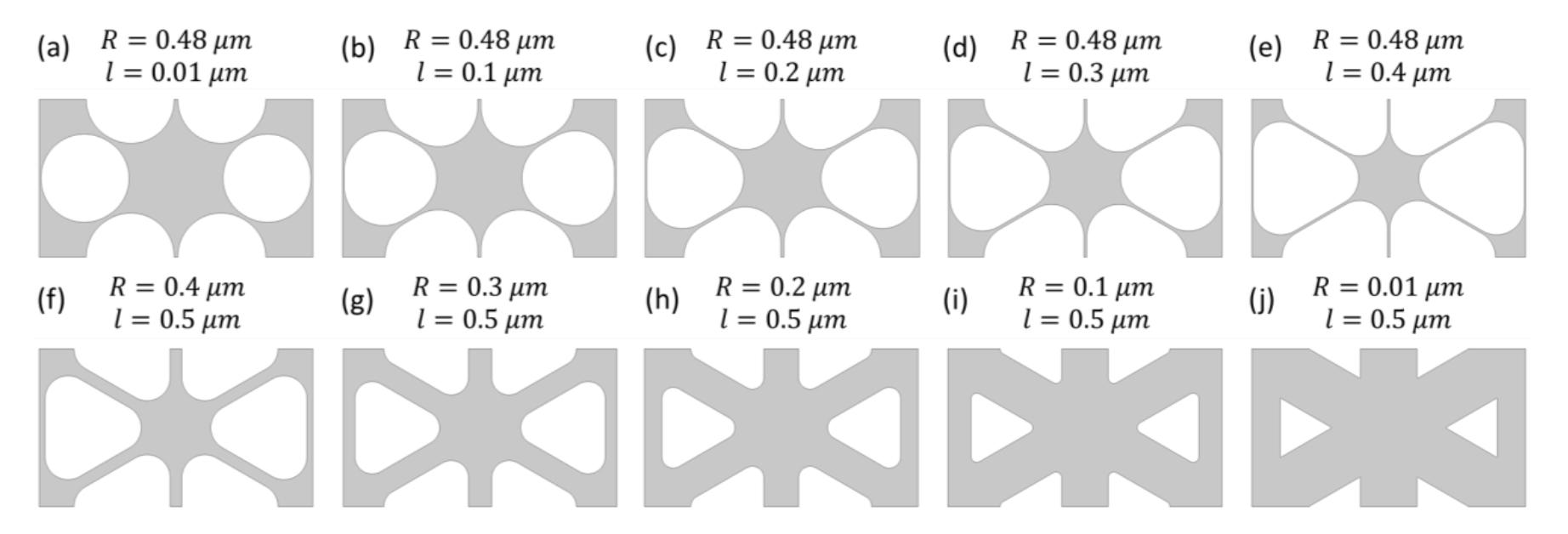}
\caption{(a-e) Change in geometry observed by increasing \textit{l} value and keeping the \textit{R} value constant. The air gap holes gradually transform from circular to triangular; (f-j) Change in geometry observed by decreasing the \textit{R} value. The air gap holes are observed to reduce in size with decrease in \textit{R} value and gradually transform from triangular holes of filleted vertices to sharp vertices.}
\label{fig2}
\end{figure}
   
\section{Results}

\begin{figure}[H]
\centering
\includegraphics[width=1\linewidth]{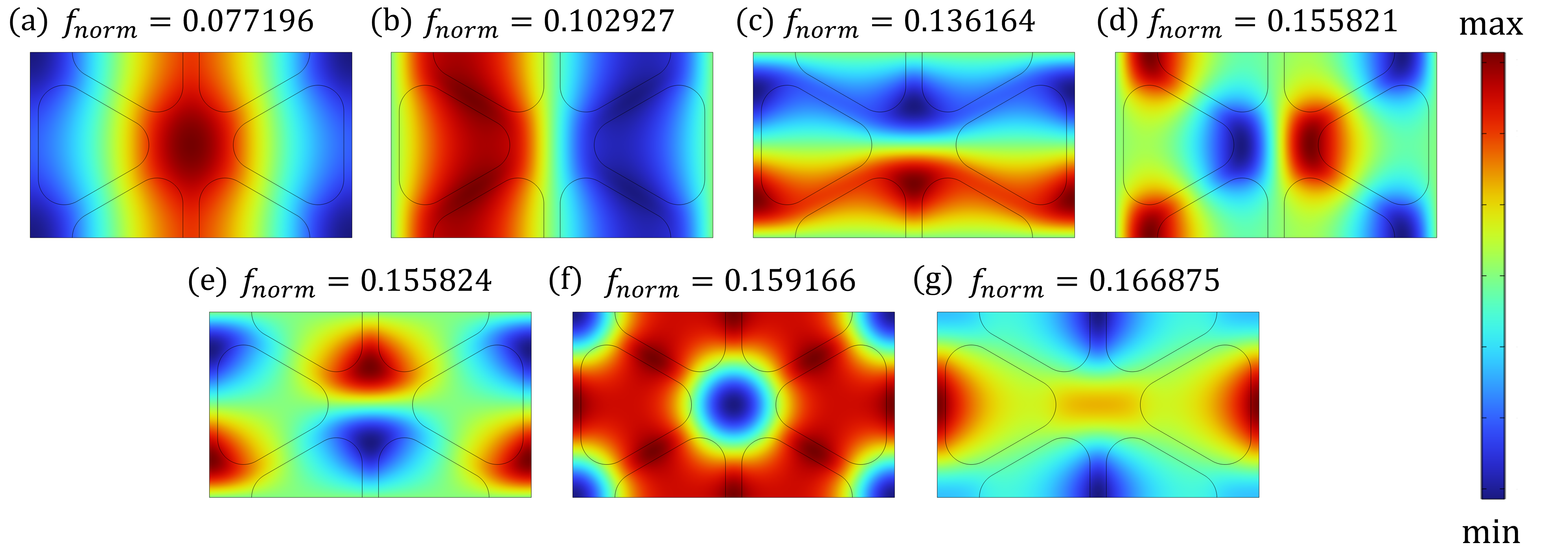}
\caption{Electric field plots showing the z-component of the initial seven eigenmodes of the unit cell with parameters $\textit{R} = 0.38\,\mu m$ and $\textit{l} = 0.5\,\mu m$. Each subplot (a–g) corresponds to a distinct normalized eigenfrequency as indicated above each plot, visualized using a blue-red colormap where red denotes regions of maximum field intensity and blue denotes minimum intensity. Within the analyzed seven eigenmodes, plots (d-e) correspond to a pair of degenerate mode with $f_{norm} \approx 0.155$. Note: The normalized frequency $f_{norm}$ is calculated by multiplying the eigenfrequency by a normalization factor $1\,\mu m/c$, where $c$ is the speed of light in vacuum ($3 \times 10^{8}\,m/s$).}
\label{fig3}
\end{figure}

\begin{figure}[H]
\centering
\includegraphics[width=1\linewidth]{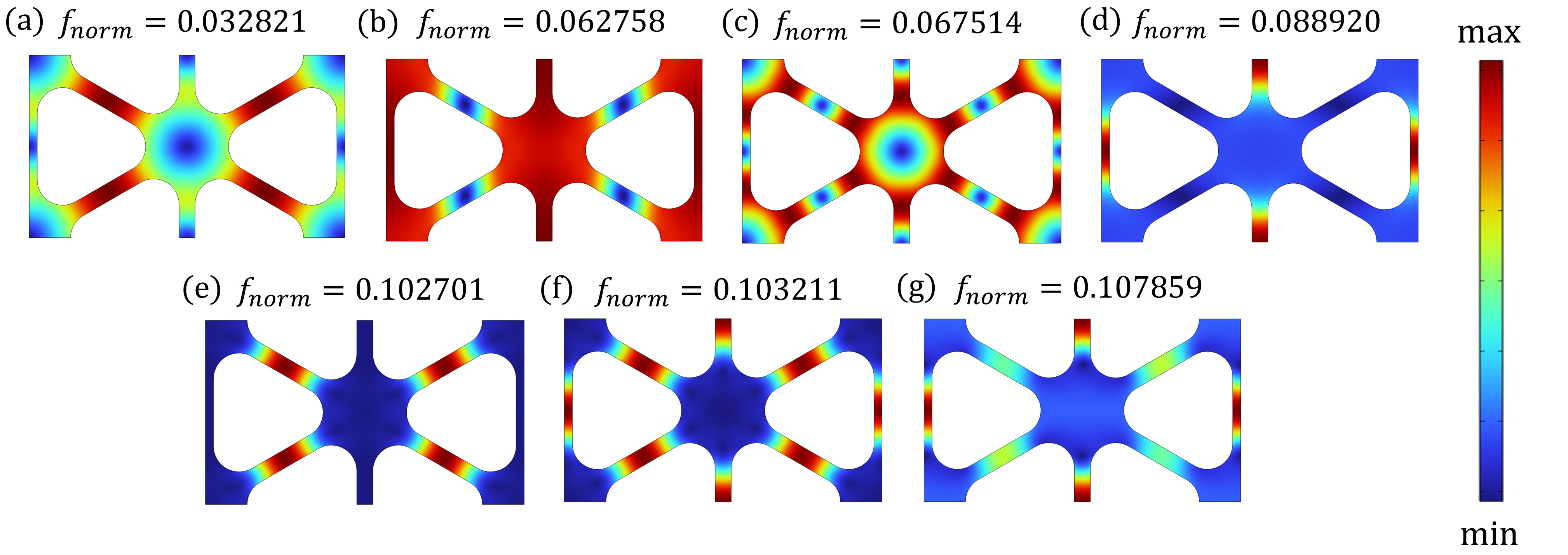}
\caption{Solid displacement plots showing the z-component of the initial seven eigenmodes of the unit cell with parameters $\textit{R} = 0.38\,\mu m$ and $\textit{l} = 0.5\,\mu m$. Each subplot (a–g) corresponds to a distinct normalized eigenfrequency as indicated above each plot, visualized using a blue-red colormap where red denotes regions of maximum field intensity and blue denotes minimum intensity. Within the analyzed seven eigenmodes, plots (e-f) correspond to a pair of degenerate mode with $f_{norm} \approx 0.103$. Note: The normalized frequency $f_{norm}$ is calculated by multiplying the eigenfrequency by a normalization factor $1\,\mu m/c$, where $c$ is the speed of sound in Silicon ($8.518.4\,m/s$).}
\label{fig4}
\end{figure}

The photonic and phononic eigenfrequencies are solved with the aforementioned geometric, material and periodic constraints. Henceforth, 2D magnitude plots are obtained for electric field (Fig.~3) demonstrating photonic wave propagation and solid displacement (Fig.~4) demonstrating phononic wave propagation for the geometry \textit{R} = 0.3 $\mu m$, \textit{l} = 0.5 $\mu m$. Here, for these plots, we considered \textit{k} = 3. To find the photonic and phononic bandgaps, the photonic and phononic eigenfrequencies are plotted as a function of k-vector. The size of the regions where photonic and phononic wave propagation are absent is identified as a bandgap (Fig.~5). Photonic bandgap is formed between photonic modes 2 and 3 (Fig.~5(a)), and phononic bandgap is formed between phononic modes 6 and 7 (Fig.~5(b)).

\begin{figure}[H]
\centering
\includegraphics[width=1\linewidth]{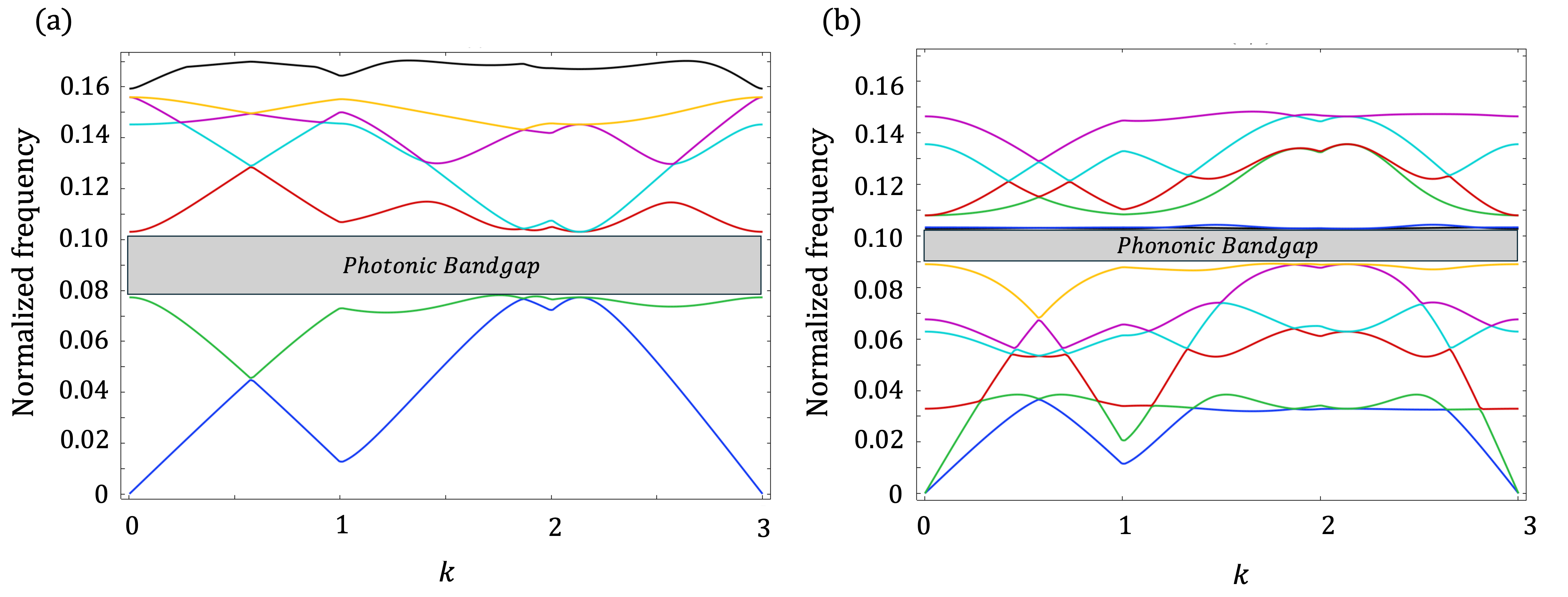}
\caption{(a) Normalized frequency of each of the initial photonic modes vs. $k$-vector for the geometry $R = 0.38\,\mu m$, $l = 0.5\,\mu m$, with the photonic bandgap highlighted in grey; (b) Normalized frequency of each of the initial phononic modes vs. $k$-vector for the geometry $R = 0.38\,\mu m$, $l = 0.5\,\mu m$, with the phononic bandgap highlighted in grey. Here, the normalized frequency is the eigenfrequency multiplied by a normalization factor $1\,\mu m/c$, where $c$ is the speed of light ($3 \times 10^{8}\,m/s$) for photonic wave propagation and the speed of sound in Silicon ($8518.4\,m/s$) for phononic wave propagation.}
\label{fig5}
\end{figure}

In order to validate the bandgaps predicted by the dispersion curves in Fig.~5, transmission loss simulations were performed along the two orthogonal axes on a finite crystal array consisting of 25 unit cells ($5$ $\times$ $5$ unit cell array). Figs.~6 and 7 present the resulting transmission spectra for electromagnetic and phononic waves, respectively. For the Transverse Electric (TE) polarization, a substantial signal attenuation is observed between normalized frequencies 0.078 and 0.102, demonstrating excellent agreement with the photonic bandgap predicted in Fig.~5(a). Conversely, the Transverse Magnetic (TM) polarization (Fig.~6(b)) does not exhibit a wide continuous bandgap, but rather distinct resonant dips. This contrast confirms the inherent polarization sensitivity of high-contrast hexagonal lattices, which favors confinement of TE modes over TM modes \cite{joannopoulos1997photonic, winn1994two}. Thus, in this study, we focus on the Transverse Electric (TE) polarization, as it supports wide photonic bandgaps in the proposed hexagonal geometry. All reported photonic bandgap values refer to TE modes. \\

Similarly, Fig.~7 illustrates phononic transmission in the lattice structure. A robust phononic bandgap is evident for longitudinal waves (Fig.~7(a)), characterized by a significant drop in transmission reaching -80 dB. This high-attenuation region spans the normalized frequency range of 0.089 to 0.102, showing excellent agreement with the predicted phononic bandgap in Fig.~5(b). Notably, the transmission spectrum exhibits a distinct V-shaped profile compared to the clear bandgap predicted for an infinite periodic lattice. This phenomenon is attributed to evanescent wave tunneling through the finite structure \cite{yang2002ultrasound}. As demonstrated in prior studies, transmission near band edges in finite crystals is governed by the exponential decay of the wave amplitude. Therefore, for finite crystal lengths, this tunneling effect results in measurable transmission at frequencies that would otherwise be forbidden in an infinite periodic structure. In the case of transverse wave transmission (Fig.~7(b)), the spectrum shows a significantly narrower and partially attenuated band compared to the longitudinal case. While a suppression of signal is observed, the presence of resonant transmission peaks within the predicted phononic bandgap region indicates that the confinement of transverse (shear) modes is less robust. This behavior is explained by the anisotropic nature of the phononic bandgap, where the transmission efficiency is strongly dependent on the polarization vector of the incident wave relative to the lattice symmetry \cite{mohammadi2008evidence}. \\

\begin{figure}[H]
\centering
\includegraphics[width=1\linewidth]{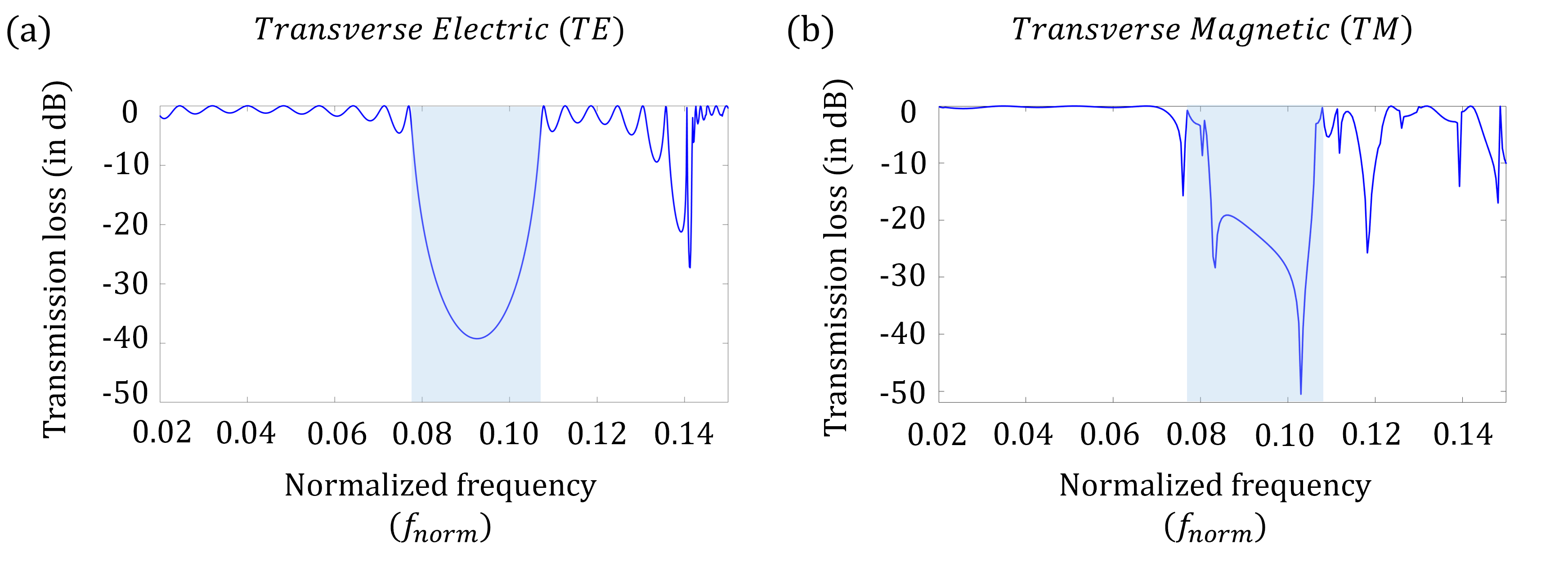}
\caption{Transmission spectrum plot with the blue region depicting photonic bandgap attenuation. (a) Transmission response for the Transverse Electric (TE) polarization (\textbf{E}-field out-of-plane), revealing a wide, complete photonic bandgap between normalized frequencies 0.078 and 0.102 with a maximum attenuation of -40 dB. (b) Transmission response for the Transverse Magnetic (TM) polarization (\textbf{E}-field in-plane), indicating the absence of a continuous bandgap.}
\label{fig6}
\end{figure}

\begin{figure}[H]
\centering
\includegraphics[width=1\linewidth]{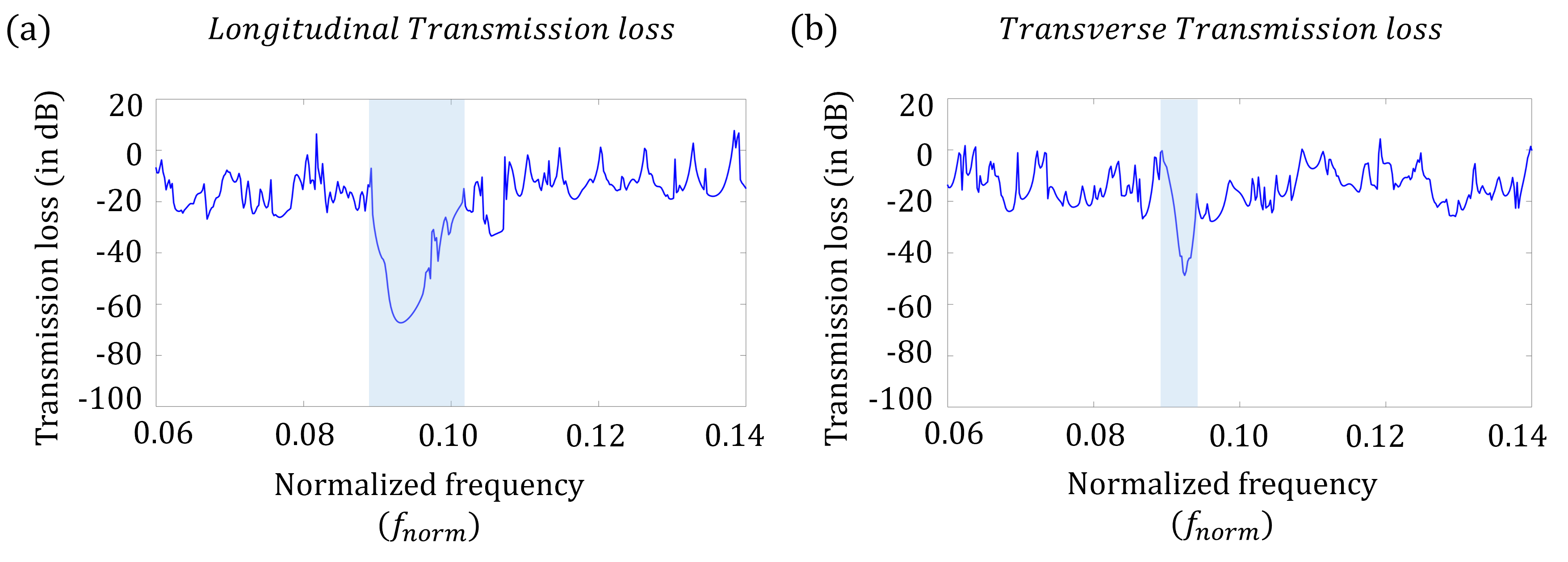}
\caption{Transmission spectrum plot with the blue region depicting phononic bandgap attenuation. (a) Transmission response for longitudinal wave excitation, displaying a deep phononic bandgap with maximum attenuation reaching -80 dB. The effective transmission gap is narrower than the infinite periodic lattice due to finite-size tunneling effects near the band edges. (b) Transverse wave transmission showing partial attenuation, confirming the anisotropic nature of the bandgap.}
\label{fig7}
\end{figure}

After confirming the formation of robust photonic and phononic bandgaps in the lattice structure through  transmission loss plots (Fig.~6 and 7), we performed independent parametric sweeps of \textit{R} and \textit{l} to study their individual geometric influences on the bandgaps. \\

By keeping \textit{a} and \textit{l} constant at $1\,\mu m$ and $0.5\,\mu m$ respectively, the normalized eigenfrequencies are recorded for different values of \textit{R}. Fig.~8(a) shows that the eigenfrequencies corresponding to both the upper and lower bounds of the photonic forbidden zone increase with \textit{R}. This is because the overall air gaps are larger for higher \textit{R}, resulting in a decreased refractive index. The eigenfrequency $f$ is related to the refractive index \textit{n} and cavity length $L$ as: 

\begin{equation}
f = \frac{c}{2nL}
\end{equation}

Since the eigenfrequency is inversely proportional to the refractive index, the eigenfrequency is higher for larger \textit{R}. However, there is a geometrical constraint that \textit{R} should be less than $0.5\,\mu m$. Hence, the maximum value of the photonic bandgap that can be achieved is $0.0616$ (normalized) for $R \to 0.5\,\mu m$. \\

Now, we study the dependence of the photonic bandgap on \textit{l}. Figs.~8(b) and 8(c) show that the eigenfrequencies corresponding to both the upper and lower bounds of the photonic forbidden zone decrease with \textit{l}. Despite the size of the air gap being higher for larger \textit{l}, leading to a reduced refractive index, the cavity length $L$ increases much more with \textit{l}. Hence, the eigenfrequency is lower for larger \textit{l}. Since the upper bound frequency decreases at a slower rate than the lower bound frequency, there exists a value of \textit{l} at which the bandgap reaches its maximum value.

\begin{figure}[H]
\centering
\includegraphics[width=1\linewidth]{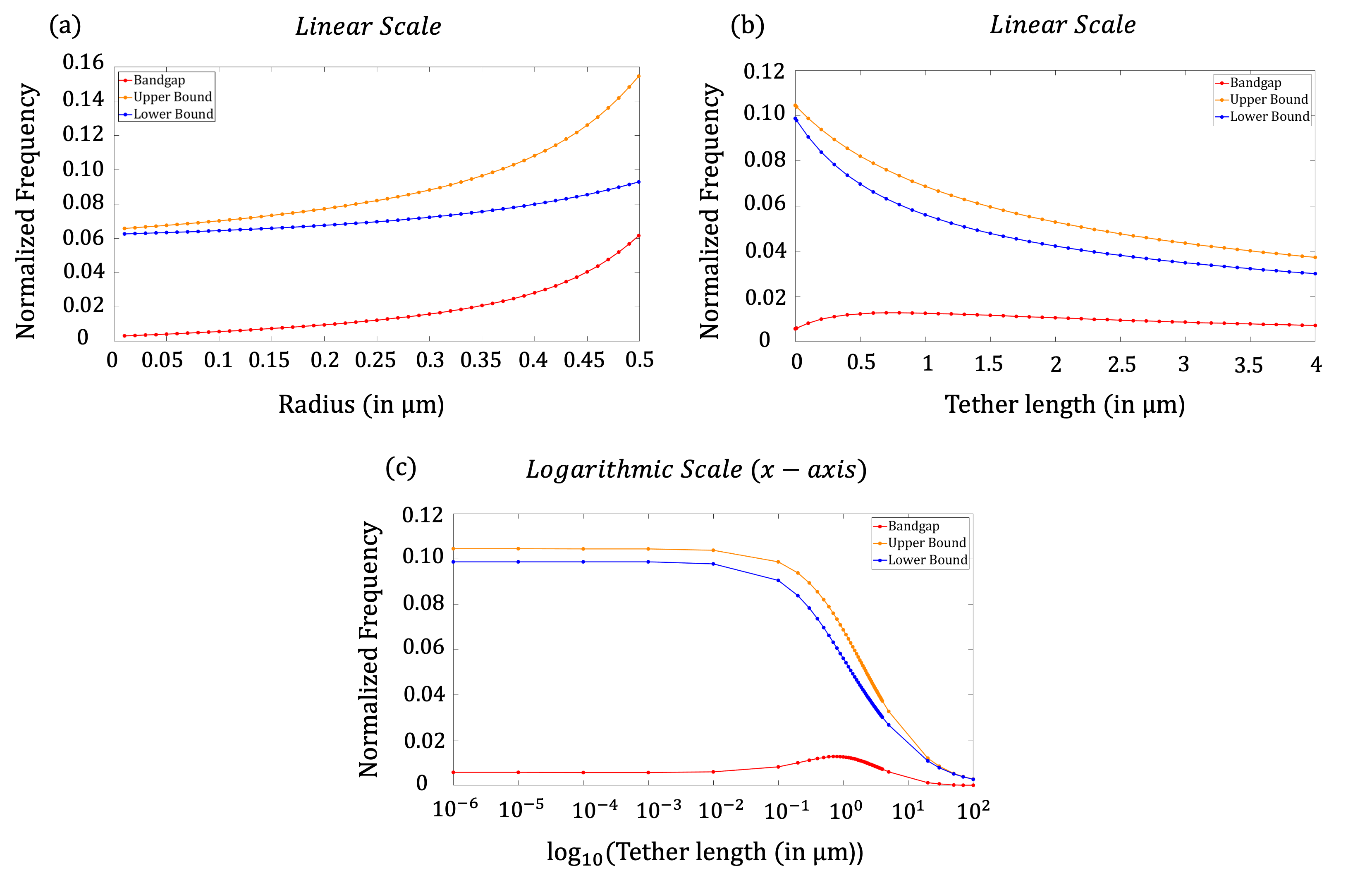}
\caption{(a) Dependence of photonic bandgap on \textit{R} for a fixed value of $l = 0.5\,\mu m$; (b) Dependence of photonic bandgap on \textit{l} for a fixed value of $R = 0.25\,\mu m$ wherein the bandgap reaches its peak value at $l = 1.4\,\mu m$}; (c) Same as (b) but the x-axis is log-scaled.
\label{fig8}
\end{figure}

By keeping $a$ and $l$ constant at $1\,\mu m$ and $0.5\,\mu m$ respectively, the normalized phononic eigenfrequencies are recorded for different values of $R$. Fig.~9(a) shows that the eigenfrequencies corresponding to both upper and lower bounds of the phononic forbidden zone decrease with $R$. The eigenfrequency $f$ is related to mass $m$ and bending stiffness $k$ as: 

\begin{equation}
f = \sqrt{\frac{k}{m}}
\end{equation}

Since the magnitude of reduction in bending stiffness is much more than the reduction in mass with an increase in air gap at larger $R$ values, the eigenfrequency decreases with increasing $R$. Also, as seen in Fig.~9(a), there is an abrupt bend at $R \sim 0.31\,\mu m$. This is because the bandgap is bounded by a different phononic mode on the lower side, whose eigenfrequency has a different decay rate with $R$. As a result of this abrupt bend in normalized frequency vs.~$R$ curve for the lower bound, the associated bandgap also reaches its peak at $R \sim 0.31\,\mu m$. Similar to the photonic bandgap, Figs.~9(b) and 9(c) also show a similar trend for the dependence of the phononic bandgap on $l$. Both upper-bound and lower-bound eigenfrequencies decrease with increasing $l$ due to the increased cavity length, but at different rates — hence imposing a critical value for $l$ to achieve an optimal bandgap. \\

\begin{figure}[H]
\centering
\includegraphics[width=1\linewidth]{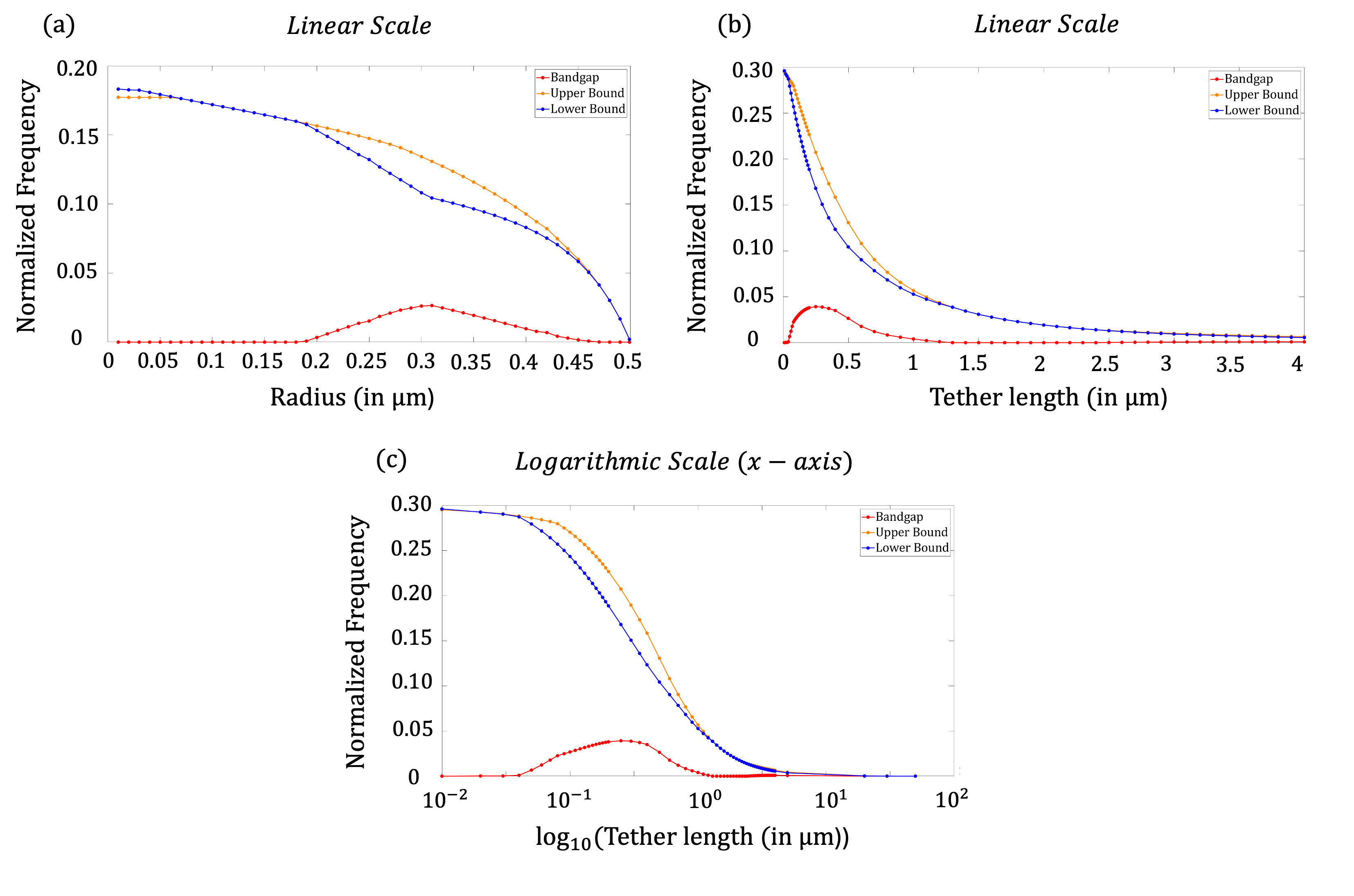}
\caption{Dependence of phononic bandgap on \textit{R} for a fixed value of \textit{l} = $0.5\,\mu m$, wherein the bandgap reaches its peak value at \textit{R} $\sim 0.31\,\mu m$; (b) Dependence of phononic bandgap on \textit{l} for a fixed value of \textit{R} = $0.31\,\mu m$ wherein the bandgap reaches its peak value at \textit{l} = $0.25\,\mu m$; (c) Same as (b) but the x-axis is log-scaled.}

\label{fig9}
\end{figure}

Having established the independent behavior of the bandgaps, we now identify the specific geometrical parameters corresponding to the coexistence of photonic and phononic bandgaps. Fig.~10 maps the normalized bandgaps against the lattice parameters to visualize this simultaneous operation window. In Fig.~10(a), the overlapping region ($0.2\,\mu m$ $< R < 0.48\,\mu m$) represents the range of radii for which the structure exhibits coexisting phoxonic bandgaps. Notably, while the photonic gap widens unrestricted with $R$, the phononic gap dictates the upper and lower bounds of this window. \\

Similarly, Fig.~10(b) further examines the stability of this simultaneous existence with respect to the tether length $l$ (at fixed $R=0.25\,\mu m$). The photonic gap remains robust across several orders of magnitude of tether length ($10^{-3}$ to $1\,\mu m$), whereas the phononic gap is strictly confined to a narrow window ($0.1\,\mu m$ $\leq l \leq 1.0\,\mu m$). Both these plots demonstrate that while the photonic gap is stable over a broad range of air-gap hole radius and tether lengths, the phononic gap imposes stricter geometric constraints. \\

\begin{figure}[H]
\centering
\includegraphics[width=1\linewidth]{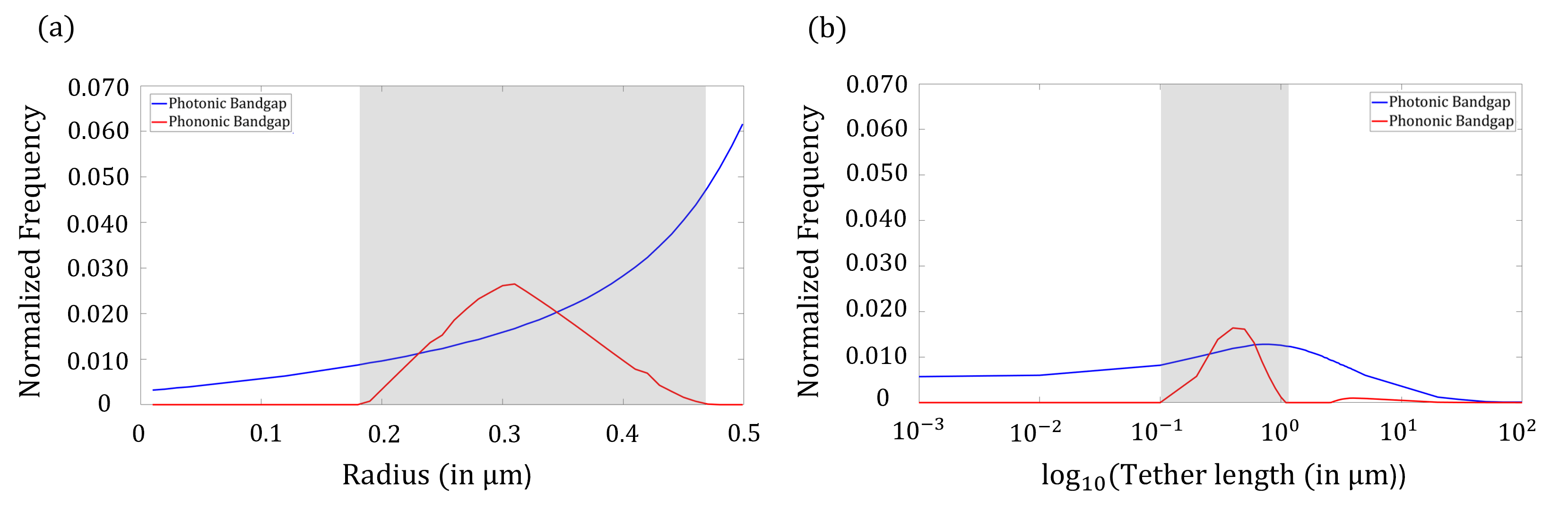}
\caption{(a) Evolution of photonic (blue) and phononic (red) bandgaps with respect to $R$ for a fixed $l$ value of $0.5\,\mu m$. (b) Evolution of bandgap width with respect to tether length $l$ (log-scale), for a fixed $R$ value of $0.25\,\mu m$ photonic and phononic bandgap. The overlapping region where both curves are non-zero (shaded in gray) defines the simultaneous operation window.}
\label{fig10}
\end{figure}

This comprehensive analysis demonstrates simultaneous photonic and phononic bandgaps in a 2D hexagonal lattice with geometry-dependent circular-to-triangular air gap transitions. By performing parametric sweeps of $R$ and $l$, we have established a robust framework for tuning dual bandgaps in silicon-based phoxonic crystals. By tuning $R$ and $l$ to critical values, we achieved a maximum relative bandgap of 49.7\% for photonics and 24.8\% for phononics, surpassing prior designs \cite{safavi2010design, pennec2010simultaneous, djafari2011band, el2010dual, ma2021dual}. Relative bandgap tunability is defined as the ratio of bandgap width to midgap frequency, expressed as a percentage, where midgap frequency is the average of the upper and lower frequency of the band. This normalization enables cross-structure comparison independent of absolute operating frequency. Tunability ranges represent the minimum and maximum achievable values across the parameter optimization space (e.g., radius variation, thickness variation, filling factor variation, etc.) reported for various designs. \\

In practical scenarios, due to design and size constraints manufacturers are often restricted to vary only one of the two variables. Hence, for such cases, the photonic and phononic bandgap tunability was evaluated under each transformation separately. For the radius variation \textit{R}, the structure supports a maximum relative photonic bandgap of 49.8\% and a relative phononic bandgap of 20.7\%. Under tether-length variation \textit{l}, the photonic and phononic bandgap widths similarly reach their respective maxima of 22.3\% and 24.8\%.

\begin{table}[H]
\centering
\label{tab1}
\begin{tabularx}{\textwidth}{|Y|Y|Y|Y|Y|}
\hline
\textbf{Structure Geometry} & \textbf{Tuning Parameter} & \textbf{Max. Photonic Tunability} & \textbf{Max. Phononic Tunability} & \textbf{Reference Work} \\
\hline\hline
\textbf{Honeycomb (Circular-to-Triangular holes)} & \textbf{Radius of air gap and tether length} & $\mathbf{0\%}$ \textbf{to} $\mathbf{49.7\%}$ & $\mathbf{0\%}$ \textbf{to} $\mathbf{24.8\%}$ & \textbf{This work} \\
\hline
Quasi-2D Cross Structure & Cross width and height & $-$ & $\sim 2.8\%$ to $7.4\%$  & \cite{safavi2010design} \\
\hline
Quasi-2D Snowflake Structure & Snowflake width and radius  & $\sim 1.2\%$ to $8.7\%$ & $\sim 0.8\%$ to $15.5\%$ & \cite{safavi2010design} \\
\hline
Square lattice (air-gap holes) & Fill factor and slab thickness & $\sim 0\%$ to $11.7\%$ & $\sim 0\%$ to $54\%$ & \cite{pennec2010simultaneous} \\
\hline
Honeycomb lattice (air-gap holes) & Fill factor and slab thickness & $\sim 0\%$ to $8.6\%$ & $\sim 0\%$ to $123\%$ & \cite{pennec2010simultaneous} \\
\hline
BN lattice (air-gap holes) & Fill factor, radius and slab thickness & $\sim 0\%$ to $35.2\%$ & $\sim 0\%$ to $54.5\%$ & \cite{pennec2010simultaneous} \\
\hline
Square lattice (nanoholes) & Fill factor, radius and slab thickness & $\sim 10.2\%$ to $17.3\%$  & $\sim 12.3\%$ to $21.3\%$  & \cite{djafari2011band} \\
\hline
BN lattice (nanoholes) & Fill factor, radius and slab thickness & $\sim 0\%$ to $36.3\%$ & $\sim 0\%$ to $36.8\%$ & \cite{djafari2011band} \\
\hline
Square lattice (Silicon pillar) & Plate thickness and height of pillar & $\sim 0\%$ to $8\%$ & $\sim 0\%$ to $42.8\%$ & \cite{el2010dual} \\
\hline
Triangular lattice (Silicon pillar) & Plate thickness and height of pillar & $\sim 0\%$ to $7.6\%$ & $\sim 0\%$ to $41.2\%$ & \cite{el2010dual} \\
\hline
Honeycomb lattice (Silicon pillar) & Plate thickness and height of pillar & $\sim 0\%$ to $6.7\%$ & $\sim 0\%$ to $21.2\%$ & \cite{el2010dual} \\
\hline
Square Lattice with Cross-Beam Topology & Thickness of plate, length of silicon lattice and length of scatterer & $\sim 5\%$ to $10.7\%$ & $\sim 40\%$ to $61.4\%$ & \cite{ma2021dual} \\
\hline
\end{tabularx}
\caption{Comparison of the tunability achieved in different Si-based 2D phoxonic geometries.}
\end{table}

To contextualize our geometric tuning framework, Table 1 compares our simulation results against other silicon-based 2D phoxonic crystal designs reported in literature. Our design demonstrates superior photonic tunability of 49.7\% using circular-to-triangular holes (versus 8.6\% for purely circular air-gap holes) while maintaining a respectable phononic bandgap tunability (upto 24.8\%). In contrast, cross-beam topologies and silicon pillars achieve higher phononic tunability (61.4\% and 41.2\%) but at the expense of photonic performance ($\sim 5$–$10.7\%$). We thus achieve the critical advantage of balanced simultaneous control of both phononic and photonic domains which is essential for integrated acousto-optic and opto-mechanical applications where neither of the domains can be compromised. \\

\section{Conclusions}

This study presents a geometrically tunable 2D silicon hexagonal lattice that simultaneously supports both photonic and phononic bandgaps through a continuous transformation of air-gap hole shape from circular to triangular. In the case of photonics, the bandgap reaches a maximum value of 0.0616 (normalized) when $R$ reaches its maximum possible value, i.e., $R \to 0.5a$. In contrast, there exists a critical value of $l$ for obtaining the maximum bandgap. For phononics, there exist critical values for both $R$ and $l$ to achieve an optimal bandgap. Our result is significant because the presented phoxonic structure supports geometric programmability using $l$ and $R$. This static programmability—the embedding of tunable filtering characteristics directly into the lattice geometry during the design phase—offers a reproducible, fabrication-friendly route to high-$Q$ microcavities \cite{safavi2010design}, on-chip Bragg filters \cite{oser2019coherency}, and co-localized photonic–phononic waveguides \cite{shin2015control} for integrated optomechanical and acousto-optic applications. Moreover, by independently varying the hole radius ($R$) and tether length ($l$), we achieve bandgap tuning up to 49.7\% for photonics and 24.8\% for phononics. This work, thus, establishes a universal geometric framework for the predictive design of phoxonic crystals in silicon, eliminating the need for complex defects, exotic materials, and cryogenic operational conditions. The proposed architecture lays the groundwork for next-generation programmable photonic–phononic devices in photonic integrated circuits \cite{fu2019phononic}, telecommunications \cite{oser2019coherency}, sensing platforms \cite{lucklum2013phoxonic}, and optomechanical frequency combs \cite{zhang2021optomechanical, de2023mechanical, wang2024optomechanical, xiao2024broadband}, where tailored phononic–photonic interactions can be harnessed with high precision and scalability.

\clearpage
    
\bibliographystyle{IEEEtran}
\bibliography{ref}

\end{document}